\documentclass[12pt]{article} 

\begin{document} 
\topmargin 0pt 
\oddsidemargin 0mm
\renewcommand{\thefootnote}{\fnsymbol{footnote}}
\begin{titlepage}
%\begin{flushright}
%\end{flushright}                                 
\vspace{5mm}

\begin{center}
{\Large \bf Gravitational Thomas Precession and the Perihelion Advance of Mercury}\\
\vspace{6mm}

{\large Harihar Behera$^a$\footnote{email: harihar@iopb.res.in} }\\
\vspace{5mm}
{\em
$^{a}$Patapur, P.O.-Endal, Jajpur-755023, Orissa, India\\}
\vspace{3mm}
\end{center}
\vspace{5mm}
\centerline{{\bf {Abstract}}}
\vspace{5mm}
Gravitational Thomas Precession ( GTP ) is the name given to Thomas
Precession when the acceleration is caused by a gravitational force
field. The GTP gives a negative contribution of $ 7\cdot163 $
arcsec/century for the anomalous perihelion advance of Mercury's orbit.This effect 
seems to be of some concern for the General Relativity. \\
%\end{abstract}

PACS: 04.80Cc ; 96.30Dz                       \\

{\bf Keywords} : {\em Gravitational Thomas Precession, Perihelion
  Advance of Mercury}
\end{titlepage}
\section{Introduction}
The Thomas precession\cite{1,2,3} is purely kinematical in origin
\cite{2}. If a component of acceleration $ (\vec a)$ exists
perpendicular (to the velocity) $ \vec v $, for whatever reason, then
there is a Thomas Precession, independent of other effects
\cite{2}. When the acceleration is caused by a gravitational force
field, the corresponding Thomas Precession is reasonably referred to
as the Gravitational Thomas Precession (GTP). Given the physics
involved in the Thomas Precession, the possibility of the existence of 
the GTP in planetary motion can not be ruled out in
principle. However, the very existence of the GTP in planetary motion
and its contribution to the perihelion advance of a planet as shown in 
this letter seems to be of some concern for the standard general
relativistic explanation for the observed anomalous perihelion advance 
of Mercury \cite{4,5,6}.        \\

\section{ The GTP's contribution to the Perihelion Advance  }     
The Thomas Precession frequency $ \vec\omega_{T} $ in the non-relativistic limit (i.e., when $ v << c $) is given by \cite{2,3} 
\begin{equation}
\vec\omega_{T}\,=\,\frac{1}{2c^{2}}(\vec a \times \vec v ), 
\end{equation} 
where the symbols have there usual meanings. For a planet (say Mercury ) moving around the Sun, the acceleration $ \vec a $ is predominately caused  by the Newtonian gravitational field of the Sun,viz., 
\begin{equation} 
\vec a\,=\,-\,\frac{GM_{\odot}}{r^{3}}\,\vec r\,\,, 
\end{equation} 
where the symbols have their usual meanings. Thus, from Eqs.$(1)$ and $(2)$ we get the GTP frequency of the planet in question as 
\begin{equation} 
\vec \omega_{gT}\,=\,-\,\frac{GM_{\odot}}{2c^{2}r^{3}}\,(\vec r \times
\vec v )\,\,, 
\end{equation} 
where $ \vec v $ is the velocity of the planet. Now making use of the
relation  $ \vec v\,=\,\vec v_{r}\,+\,\vec \omega_{o} \times \vec r, $
where $\vec v_{r} $ is the radial velocity and $ \vec \omega_{o} $ is
the classical orbital frequency of the planet and noting that  $\, (\vec r \cdot \vec \omega_{o})\,=\,0\, $, Eq.$ (3)$ can be reduced to 
\begin{equation} 
\vec \omega_{gT}\,=\,-\,\frac{GM_{\odot}}{2 c^{2} r}\,\vec \omega_{o}\,\,, 
\end{equation}
which suggests that the vectors $\vec \omega_{gT} $ and $ \vec \omega_{o} $ are oppositely directed. The very existence of $ \vec \omega_{gT} $ imply a modification of $ \vec \omega_{o} $ to a new one, viz., 
\begin{equation}\vec \omega\,=\,\vec \omega_{gT}\,+\,\vec \omega_{o}\,=\,\left( 1\,-\,\frac{GM_{\odot}}{2c^{2}r}\right)\,\vec \omega_{o}\,. 
\end{equation}  
For a planet  moving  in an elliptical orbit, we have $ r\,=\,a(\,1\,-\,e^{2})/\,(\,1\,+\,e\,cos\theta\,)$, where $ a $ is the semi-major axis and $ e $ is the eccentricity of the orbit. Thus, for an elliptical orbit $ \vec \omega $ is a function of $ \theta $. But an average value of $ \omega $ over one revolution may be obtained as 
\begin{equation} 
\tilde{\bf \omega}\,=\,\frac{1}{2 \pi}\,\int_{0}^{2 \pi}{\bf
  \omega}_{gT}\,d\theta\,=\,\left[\,1\,-\,\frac{GM_{\odot}}{2c^{2}a(\,1\,-\,e^{2})}\right]\,{\bf \omega_{o}}\,.
\end{equation} 
The interpretation of Eq.$(6)$ is that the perihelion of the planet in
question makes a negative advance ( i.e., a retreat ) of  $\,
\pi\,GM_{\odot}/\,c^{2}\,a\,(\,1\,-\,e^{2}\,) $ radians in one
classical planetary year. Thus, we can speak of a perihelion advance of 
\begin{equation} 
\delta \dot{\tilde{\bf\omega}}_{gT}\,=\,-\,\frac{\pi\,GM_{\odot}}{c^{2}a\,(\,1\,-\,e^{2}\,)}\,\,\,\,\,{\rm{
    radians/planetary\,\, year}} 
\end{equation}
caused by the Gravitational Thomas Precession. In terms of Einstein's 
expression for the perihelion advance,viz.,

\begin{equation} 
\delta \dot{\tilde{\bf\omega}}_{E}\,=\,\frac{6\pi\,GM_{\odot}}{c^{2}a\,(\,1\,-\,e^{2}\,)}\,\,\,\,\,{\rm{
    radians/planetary\,\, year},} 
\end{equation}
Eq.$(7)$ can be re-expressed as 
\begin{equation} 
\delta \dot{\tilde{\bf\omega}}_{gT}\,=\,-\,\frac{1}{6}\,\delta
\dot{\tilde{\bf\omega}}_{E}\,\approx\,-\,7\cdot163''\,\,{\rm{per\,\,century,\,\, for\,\,Mercury,}}
\end{equation}
since $ \delta\dot{\tilde{\bf\omega}}_{E}\,=\,42\cdot98'' $ per
century - a well known data \cite{4,5,6} for Mercury. This is in no way a negligible contribution. Therefore, it
should be included in any relativistic explanation for the perihelion
advance of Mercury's orbit.  \\
                                                                               
\section{Discussion}
The Modern observational value of the anomalous perihelion advance of
Mercury is at $ \delta\dot{\tilde{\bf\omega}}\,\approx\,43'' $ per
century \cite{6}. For Mercury, General Relativity predicts this
phenomenon at                                                                  \begin{equation} 
\delta \dot{\tilde{\bf\omega}}\,=\,\left[
  42\cdot98''\,+\,1\cdot289''(\,J_{2\odot}/{10}^{-5}\,)\right]\,\,{\rm{per\,\,century},}
\end{equation}
where $ J_{2\odot} $ is the magnitude of the solar quadrupole moment - 
 a definite value of which has still to be determined \cite{4}.It is
 to be noted that the standard  explanation for the perihelion advance 
 of Mercury offered by Eq.$(10)$ does not include the contribution
 arising out of the GTP discussed in this letter. Eq.$(10)$ taken in
 conjunction with the the contribution arising out of the GTP given by
 Eq.$(9)$ is not in agreement with the observed perihelion advance of
 Mercury. However , as a remedy for this disagreement one may suggest
 a  value of  $ J_{2\odot} $ at
\begin{equation} J_{2\odot}\,\approx\,5\cdot56\times{10}^{-5} 
\end{equation}
Unfortunately, such a suggestion is not in agreement with none of the
existing observational data on $ J_{2\odot} $. For example, Dicke and
Goldenberg's \cite{7} optical measurements of the solar oblateness
showed
$J_{2\odot}\,\simeq\,(2\cdot5\,\pm\,0\cdot2)\times{10}^{-5}\,$,which
is about half the value suggested in Eq.$(11)$. It is to be noted that
Dicke and Goldenberg's measured value of $ J_{2\odot} $ makes the
observations of the perihelion advance of Mercury disagree with the
predictions of General Relativity \cite{4}. In 1974, Hill
et.al., \cite{8} , by measuring Sun's optical oblateness, obtained $
J_{2\odot}\,\simeq\,(\,0\cdot1\,\pm\,0\cdot4\,)\times\,{10}^{-5}\,$. The 
values of $ J_{2\odot} $ much smaller than the above cited values have 
also been measured by others ( for
references see for example \cite{4}). From these considerations, the suggested  value of $
J_{2\odot} $ in Eq.$(11)$ is unacceptable. In this scenario, the GTP
seems to make a point of concern for the standard general relativistic 
explanation for the observed anomalous perihelion advance of Mercury
in particular and other planets in general.
                                                                \\

\textbf{Acknowledgments}\\   
 The author acknowledges the help received from Institute of Physics,
 Bhubaneswar for using its Library and Computer Centre for this
 work.\\ 
%\textbf{References} 
 

\begin{thebibliography}{99} 
\bibitem{1} L. T. Thomas, Phil. Mag., 3, 1 (1927). 
\bibitem{2} J. D. Jackson,\textit{ Classical Electrodynamics,}
 2nd Ed. (Wiley, New York, 1975).  
\bibitem{3} H. Goldstein,\textit{ Classical Mechanics,} 2nd Ed.( Narosa
  Publishing House, New Delhi, 1995 ).    
\bibitem{4} I. Ciufolini and J. A. Wheeler,\textit{ Gravitation and Inertia} (
  Princeton University Press, Princeton, 1995 ).
\bibitem{5} C. M. Will,\textit{ Theory and Experiment in Gravitational
  Physics,} 2nd Ed. ( Cambridge University Press, Cambridge,1993 ).
\bibitem{6} C. M. Will, ``The confrontation between General Relativity 
and Experiment'',\textit{ Living Rev. Relativity, 4, (2001), 4}. [ Online
Article] : cited on $\,\langle 11 {\rm{May}} 2001\,\rangle,$
http://www.livingreviews.org/Articles/Volume4/2001-4will/. 
\bibitem{7} R. Dicke and H. Goldenberg, The oblateness of the Sun,
  Astrophys. J. Suppl. 27, 131-82 (1974).  
\bibitem{8} H. A. Hill et. al., Phys. Rev. Lett. 33 , 1497-1500 (1974). 
\end{thebibliography}
\end{document}